\documentstyle[prb,aps,epsf]{revtex}
\draft

\begin{document}


\title{Numerical study on Anderson transitions in three-dimensional disordered systems in
random  magnetic fields}
\author{Tohru Kawarabayashi\footnote{Present Address: Faculty of
Science, Toho University, Miyama 2-2-1, Funabashi, Chiba 274, Japan}}
\address{Institute for Solid State Physics, University of Tokyo,
Roppongi, Minato-ku, Tokyo 106, Japan}
\author{Bernhard Kramer}
\address{I. Institut f\"{u}r Theoretische Physik, Universit\"{a}t Hamburg,
Jungiusstra{\ss}e 9, D-20355 Hamburg, Germany}
\author{Tomi Ohtsuki}
\address{Department of Physics, Sophia University,
Kioi-cho 7-1, Chiyoda-ku, Tokyo 102, Japan}

\date{\today}
\maketitle

\begin{abstract}
The Anderson transitions 
in a random magnetic field in three dimensions are investigated numerically.
The critical behavior near the transition point is analyzed 
in detail by means of the transfer matrix method with high accuracy 
for systems both  with and without an additional random scalar potential. 
We find the critical exponent $\nu$ for the localization
length to be $1.45 \pm 0.09$ with a strong random scalar potential.
Without it, the exponent is smaller but increases with the system sizes
and extrapolates to the above value within the error bars. These results
support the conventional classification of universality classes due to
symmetry. Fractal dimensionality of the wave function at the critical 
point is also estimated by the equation-of-motion method.
\end{abstract}


\widetext

\section{Introduction}

Since the pioneering work by Anderson\cite{Anderson},
the metal-insulator transition driven by disorder, which is called
the Anderson transition(AT), has attracted much attention for
many years\cite{Wegner,Hikami,LR,KM}.
The critical behavior of the AT is conventionally classified, depending
on the symmetry of hamiltonians, into
three universality classes: the orthogonal, the unitary and the
symplectic classes\cite{Dyson}.
Systems invariant under spin rotation as well as time reversal form
the orthogonal class. The unitary class is characterized by the
absence of the time reversal symmetry. Systems invariant under time reversal
but having no spin rotation symmetry belong to the symplectic class.

In the last decade, there has been considerable progress in 
the numerical study of the AT in three dimensions(3D) 
by the finite-size scaling analysis 
for quasi-1D systems \cite{MK}. 
In the early stage,  
it was not easy to confirm numerically 
for the 3D orthogonal class that 
the critical exponent is 
insensitive to the choice of the probability distribution 
of random potential \cite{KBMS}.
This discrepancy in exponents for different distributions of 
random potential has been removed by improving the accuracy 
of numerical calculations \cite{MacKinnon} 
and by taking into account the 
corrections to scaling \cite{SO98}.
With such a high-accuracy analysis, it has been concluded that 
the critical exponent for the orthogonal system can be 
distinguished from that for the unitary system\cite{SO}.  
These recent developments confirm the universality of
critical exponents as well as the validity of 
the conventional classification of universality classes in AT.
It should be noted, however, that in most cases, such analyses 
have been restricted to the AT near the band center 
in the presence of a 
random scalar potential, where 
the scaling analysis works fairly well.
In contrast, for the AT away from the band center, 
no systematic scaling behavior has been observed\cite{KBMS,SK}.

The AT in a magnetic field has been studied extensively, mainly in
connection with
the quantum Hall effect\cite{Hajdu,Huckestein}.
Accordingly, in most cases, the magnetic field was
assumed to be uniform in space and the disorder was 
introduced by a random scalar potential. 
On the other hand, in recent years, there has also been considerable
interest in the transport properties of a system  subject to a 
spatially  random  magnetic field.
The random magnetic field introduces randomness
as well as the absence of invariance under time reversal in a system.
In fact, it has been shown\cite{OOK} that in 3D the AT occurs in the presence of
the random magnetic field and without a random scalar potential. 

The AT in a random magnetic field is driven by the coherent
scattering due to a fluctuating vector potential. 
A nontrivial feature of 
this coherent scattering by a fluctuating vector potential
has been pointed out \cite{AI} in a theory of strongly correlated 
spin systems.
Much work has also been done on
transport properties in 2D in a random magnetic field\cite{Furusaki}, in particular
in connection with the theory of the fractional quantum Hall effect \cite{HLR}
in a high magnetic field.
It is thus an important issue to understand how the effect of
coherent scattering in a strongly fluctuating random vector potential
will show up in the AT.

The magnetic field breaks the time reversal symmetry and
thus all systems in the magnetic field should belong to the unitary class.
In fact, it has been demonstrated numerically in 3D\cite{HKO,KOH} 
that in the presence of
a random scalar potential, the
critical exponent takes a universal value,
irrespective of whether the magnetic field
is uniform or random. The AT with a random 
potential and in a uniform magnetic field 
has been re-analyzed recently and the critical exponent
for the localization
length has been determined to be $1.43 \pm 0.06$\cite{SO}.

The AT, in 3D, in the presence of
a random vector potential and without
a random scalar potential, has also been
investigated based on the finite-size scaling. 
The data suggested \cite{OOK} that
the mobility edge is very close to the band edge. The exponent for
the localization length has been estimated to be
$\nu \approx 1$ \cite{OOK} which is
considerably smaller than that in the case with an additional
random scalar
potential and in a uniform magnetic field.
This seemed to indicate that  in 3D the AT driven solely by a random
vector potential might exhibit critical behavior different from that observed in
other unitary systems, for example systems having additional
random scalar potential.
Apparently, this questions the validity of the conventional
classification of universality classes in AT.
On the other hand, 
it should be recalled that the finite-size scaling analysis 
did not work for the AT near the effective band edge\cite{KBMS,SK}.
It is thus important to
re-examine the applicability of the scaling ansatz to the AT 
driven solely by the random magnetic field in which  the mobility 
edge lies quite  close to the band edge.

In this paper, we report on  a high-precision numerical finite-size scaling 
analysis for the AT in the random magnetic field. 
In order to clarify the origin of the above mentioned discrepancy 
between the critical exponent of the
AT far away from the band center induced solely by randomness
in a vector potential and the exponent obtained for other unitary
systems, we have considered systems both with and without 
an additional random potential. 
We also evaluate the fractal 
dimension of the wave functions at the critical point based on 
the equation-of-motion method.

The paper is organized as follows. In the next section, 
the hamiltonian which we adopt is introduced. The finite-size scaling
study on the critical phenomena is  
presented in section 3. In section 4, the fractal 
dimensionality of the wave function is discussed 
by  means of the equation-of-motion method. 
Section 5 is devoted to  summary and discussion.

\section{Model}

The model is defined
by the Hamiltonian \cite{OOK}
\begin{equation}
 H = V \sum_{<i,j>} \exp ({\rm i}\theta_{i,j}) C_i^{\dagger}C_j +
     \sum_i \varepsilon_i C_i^{\dagger}C_i ,
\end{equation}
where $C_i^{\dagger}(C_i)$ denotes the creation(annihilation)
operator of an
electron at the site $i$ of a 3D cubic lattice.
Energies $\{ \varepsilon_i\}$ denote
the random scalar potential distributed independently and
uniformly in the range
$[-W/2, W/2]$. The Peierls phase factors
$\exp ({\rm i}\theta_{i,j})$
describe a random vector potential or
magnetic field.
We confine ourselves to
phases $\{\theta_{i,j} \}$ which
are distributed independently and uniformly
in $[ -\pi ,\pi ]$.
The hopping amplitude $t$ is assumed to be the energy unit, $V=1$.
The phases $\{ \theta_{i,j} \}$ are related to the magnetic flux, 
for example, as 
\begin{equation}
 \theta_{i,i+\hat{x}} + \theta_{i+\hat{x},i+\hat{x}+\hat{y}} +
 \theta_{i+\hat{x}+\hat{y},i+\hat{y}} +\theta{i+\hat{y},i} = -2\pi
 \phi_i/ \phi_0 ,
\end{equation}
where $\phi_i$ and $\phi_0=hc/|e|$ denote the magnetic flux through the 
plaquette $(i,i+\hat{x}, i+\hat{x}+\hat{y},i+\hat{y})$ and the 
unit flux, respectively. Here $\hat{x}(\hat{y})$ stands for the 
unit vector in the $x(y)$-direction. 
Note that in the present system, the condition that the magnetic 
flux through a closed surface is zero is satisfied.

\section{Finite-Size Scaling Study}

We consider
quasi-1D systems with cross section
$M \times M$ \cite{MK,MacKinnon}.
The Schr\"{o}dinger equation $H \psi = E \psi$
in such a bar-shaped system
can be rewritten using transfer matrices $T_n(2M^2 \times 2M^2)$
\begin{equation}
  \left(   \begin{array}{c}
               \psi_{n+1} \\
               \psi_n
           \end{array} \right)
 = T_n \left(   \begin{array}{c}
               \psi_{n} \\
               \psi_{n-1}
           \end{array} \right) ,\quad
   T_n = \left(   \begin{array}{ll}
               E-H_n  & -I \\
                I    &  0
           \end{array} \right)
\end{equation}
($n=1,2,\ldots$)
where $\psi_n$ and $H_n$ denote the set of coefficients of
the state $\psi$
and the Hamiltonian of the $n-$th slice, respectively.
The identity matrix is denoted by $I$.
The off-diagonal parts of the
transfer matrix $T_n$ can be expressed by the identity matrix
because the phases in the transfer-direction can be
removed by a gauge transformation \cite{OOK}.
The logarithms of the eigenvalues of the limiting matrix $T$
\begin{equation}
 T \equiv \lim_{n \rightarrow \infty} [(\prod_{i=1}^n T_i)^{\dagger}
 (\prod_{i=1}^n T_i)]^{1/2n}
\end{equation}
are called the Lyapunov exponents.
The smallest Lyapunov exponent $\lambda_M$ along the bar is estimated
by a technique which uses the product of these transfer
matrices \cite{KM,MK}.
The relative accuracies for the smallest Lyapunov exponents
achieved here is $0.2\%$ for $M \le 10$ and
$0.25\% \sim 0.3\%$ for $M=12$.
The localization length $\xi_M$ along the bar is given by the inverse of
the smallest Lyapunov exponent, $\xi_M =1/ \lambda_M$.

The assumption of one-parameter scaling for
the renormalized localization length $\Lambda_M \equiv \xi_M /M$
implies
\begin{equation}
 \Lambda_M = f(\xi / M),
\end{equation}
where $\xi=\xi(E,W)$ is the relevant length scale in the limit
$M \rightarrow \infty$\cite{MK}.
Near the mobility edge $E_c(W)$, $\xi$ diverges with an exponent $\nu$
as $\xi \sim x^{-\nu}$ with
$x=(E-E_c)/E_c$. If the transition is driven by the disorder
$W$ at a constant energy,
$x=(W_c-W)/W_c$.
At the mobility edge, $\Lambda_M$ becomes
scale-invariant. The quantity $\Lambda_M$ is a smooth
function of $E$ and $W$, and we can expand it
as a function of $x$ as
\begin{eqnarray}
 \Lambda_M &=& \Lambda_c +
 \sum_{n=1}^{\infty} A_n (M^{1/\nu}x)^n .
 \label{fitcur}
\end{eqnarray}
By fitting our data to the above function,
we can determine the critical exponent $\nu$ and the mobility edge
accurately.
In practice, we truncated the series (\ref{fitcur}) 
at the third order$(n=3)$ and used
the standard $\chi^2$-fitting procedure\cite{For}.
The error bars are estimated by using the Hessian matrix
and the confidence interval is chosen to be
$95.4\%$.

For the the transition at the band
center in the presence of a strong random 
scalar potential, a clear scaling has been observed for
presently achievable sizes, $6 \leq M \leq 12$. 
In fact, all the data (84 points) 
for $M=6,8,10$, and $12$ in the range $ 17.8 \leq W \leq 19.8$ 
can be successfully fitted by the fitting function (\ref{fitcur})
up to the 3rd order, which has six fitting parameters including 
the critical point and the critical exponent.
We have estimated the critical disorder and the exponent $\nu$
to be $W_c = 18.80 \pm 0.04$ and $\nu = 1.45 \pm 0.09$ \cite{KKO}.
The renormalized localization length $\Lambda_c$ at the critical
point is $0.558 \pm 0.003$. 
The error bars of these estimations are at least a factor
of 3 smaller than those of the previous
estimates \cite{HKO}.

In contrast, in the absence of the random scalar potential ($W=0$) 
or in the presence of an additional weak random scalar potential
$(W=1)$, for which the critical point lies near the band edge, 
we have found \cite{KKO}
that the correction to scaling is not negligible. 
Near the band edge, the density of states
changes rapidly as a function of energy.
We have thus performed high-accuracy transfer matrix calculations 
for narrower energy range $|E-E_c| \leq 0.025$ 
around the critical point for $W=0$ and
$W=1$ \cite{KKO}.
In both cases($W=0$ and $W=1$), we have found that the estimation of  
the critical exponent tends to increase with the system-sizes. 
In order to extrapolate the critical exponent for $W=0$, we have made 
calculations for larger system sizes $M=14$ and $M=16$.
Here we show, in Table I,  the summary of the results for $W=0$ obtained
by the fittings with different sizes  up to $M=16$.
The relative accuracy in $\xi_M^{-1}$ achieved for $M=14$ and $M=16$ is
1\% for each sample
and 7 and 5 realizations of random phases are considered, respectively.
The scaling regime is assumed to be $[4.39, 4.44]$ as in ref. \ref{KKO}.
It is clear that the critical point exists around $E_c \approx 4.41$
(see figure 1).
In Table 1, we can see that 
the exponent $\nu$ tends to increase with the system-sizes and
is likely to saturate around $\nu \sim 1.48$.
Within the error bars, estimated values of $\nu$ for $M\geq 12$
are consistent with $1.45\pm 0.09$ obtained
for the band center as well as $1.43 \pm 0.06$ estimated
in the uniform magnetic field \cite{SO}.
No evidence has been found for $\nu \approx 1$ which was suggested by
calculations with low accuracy \cite{OOK}.
The present results support the universality of the critical exponent
in the unitary systems.
The positions of the critical points and the values of $\Lambda_c$
estimated with different combinations
of system-sizes are fluctuating for $M \geq 12$ (Table I).
The value of $\Lambda_c = 0.558 \pm 0.003$ at the band center seems to
lie inside the
range of this fluctuation. Conventionally, the value of $\Lambda_c$
is also expected to be universal in unitary systems.
Our results obtained here seem to be consistent with this universality of
$\Lambda_c$.

The mobility edge trajectory in the presence of the 
random magnetic field is shown in figure 2.
Each  critical point
(mobility edge) is estimated based on numerical data by the transfer
matrix method with
$M=6 \sim 10$. It should be noted that
there exist extended states for energies
larger than the critical energy $E_c \approx 4.41$ for $W=0$.
This type of reentrant phenomena in
the energy-disorder plane
has been commonly observed for systems with the uniform distribution
of random scalar potential \cite{BSK,DBZK}.
It is interpreted \cite{BSK} that
the enhancement of extended states for a weak additional random scalar
potential
is due to the enhancement of density of states at that energy regime.

\section{Equation-of-Motion Method}
We now turn our attention to the properties of wave
function just at the AT in random
magnetic fields.
It is well known that at the AT,
the wave function shows multifractal structure\cite{fractal} which
leads to the scale invariant behavior of conductance
distributions\cite{shapiro,markos,SO} and the energy level
statistics.\cite{SSSLS,OO,HS_3DO,Evangelou_3DO,ZK,BSZK,SZ,KOSO}

The direct way to investigate the wave functions is to
diagonalize the Hamiltonian.
This, however, is numerically very intensive.
Instead, we calculate here the time evolution of wave
packets to extract the information of fractal dimension.
We first prepare the initial wave packet $|0\rangle$
close to AT by diagonalizing a small cluster located
at the center of the system.
The time evolution of the state at time $t$ is then
obtained by
\[
|t+\Delta t\rangle =U(\Delta t)
|t\rangle
\]
where $U(\Delta t)$
is the time evolution operator.
In order to perform effectively the numerical calculation,
we approximate $U(\Delta t)$ by the products of exponential
operators as
\begin{equation}
U(\Delta t) ={\rm e}^{-{\rm i}H\Delta t/\hbar}
= U_2(p\Delta t) U_2((1-2p)\Delta t) U_2(p\Delta t) 
 +{\rm O}(\Delta t^5)
\end{equation}
with $p=(2-2^{1/3})^{-1}$ and
\begin{equation}
U_2(\Delta t) \equiv {\rm e}^{-{\rm i}H_1\Delta t/2\hbar}\cdots
{\rm e}^{-{\rm i}H_{q-1}\Delta t/2\hbar}
{\rm e}^{-{\rm i}H_q\Delta t/\hbar}{\rm e}^{-{\rm i}H_{q-1}\Delta t/2\hbar}
\cdots{\rm e}^{-{\rm i}H_1\Delta t/2\hbar} , 
\end{equation}
where $H_1,\cdots,H_q$ are decomposition of the
original Hamiltonian $H=\sum_i H_i$ which are simple enough
to diagonalize analytically.\cite{Suzuki}

The square displacement of the wave packets is defined by
\[
r^2(t)=\langle t| r^2|t \rangle .
\]
In metallic phase, $r^2(t)$ is proportional to $D t$ where
$D$ is the diffusion coefficient.
In the insulating phase, it saturates to the square of localization
length, $\xi^2$.
At AT, the anomalous diffusion\cite{shapiro2,OK}
\[
r^2(t)\sim t^{2/d}=t^{2/3}
\] 
is expected.
The fractal dimension $D_2$ is estimated from the
autocorrelation function
\[
C(t) =\frac{1}{t}\int_0^t {\rm d}t'
|\langle t'|0\rangle|^2
\]
where $C(t)$ is expected to decay as \cite{BHS}
\[
C(t)\sim t^{-D_2/d}.
\]
In Fig. 3, we show the results of $C(t)$
for the transition at the center of the band in the
presence of a strong random scalar potential($W=18.8V$).
By diagonalizing a small cluster of $7\times 7\times 7$ located at the
the center of the system,
we follow the time evolution of wave
packets in $101\times 101\times 101$ systems.
Geometric average of $C(t)$ over 10 random field and
potential configurations are performed.
By fitting the data for $t>40 \hbar/V$,
the fractal dimensionality $D_2$ is estimated to be
\[
D_2 = 1.52\pm 0.18
\]
considerably smaller than the space dimension $d=3$.
The above value is consistent with the estimate of 3D
system at AT in a strong uniform magnetic field.\cite{OK,HK}

\section{Discussions}

In summary, we have investigated in detail 
the AT in a random magnetic field
based on the transfer matrix method with 
considerably high accuracy. 
In particular, whether or not 
the AT driven solely by the random vector 
potential ($W=0$) exhibits
different critical behavior from other unitary systems has been 
discussed.  
In order to clarify the above point, 
we have performed the scaling analysis for the three
critical points, namely $E=0$, $W=0$, and $W=1$ (figure 2).
For the transition at the band center ($E=0$) in the presence of 
a strong additional random potential, a clear scaling behavior 
has been observed and the exponent $\nu$ has been estimated to be 
$1.45 \pm 0.09$. This coincides with the value obtained for a unitary 
system in a uniform magnetic field \cite{SO}.
It has been found, on the other hand, 
that the correction to scaling is not negligible
in the presently achievable sizes for
the transitions near the band edge($W=1$ and $W=0$). 
The exponents estimated for $W=0$
by larger system sizes are consistent with
those obtained for other unitary systems within the error bars.
From the size dependence of $\nu$, in contrast to the 
suggestion in ref.\ref{OOK}, no evidence has been found for
$\nu \approx 1$. 
These results indicate the universality of $\nu$ in the unitary 
class and hence support the conventional classification of the AT by
universality classes due to symmetry.

The mobility edge trajectory has been also obtained in the
presence of the random magnetic field. It's qualitative shape
turns out to be similar to those obtained for other systems
with the uniform distribution of random scalar potential.

We have also studied the diffusion of electrons 
at the AT in the presence of a random magnetic field.
By solving the time-dependent Schr\"{o}dinger equation 
numerically, we examine the time evolution of 
wave packets at the AT. From the asymptotic behavior 
of the autocorrelation function, we have extracted the 
fractal dimensionality of the critical wave function 
at the band center.

\section*{Acknowledgments}

The authors thank M. Batsch, A. MacKinnon, I. Zharekeshev
and K. Slevin
for valuable discussions.
The numerical calculations were performed on a FACOM VPP500 of
Institute for Solid State Physics, University of Tokyo and
in computer facilities of I. Institut f\"{u}r Theoretische Physik,
Universit\"{a}t Hamburg.
This work was supported in part by the EU-project FHRX-CT96-0042
and by the Deutsche Forschungsgemeinschaft
via Project Kr627/10 and the Graduiertenkolleg
``Nanostrukturierte Festk\"{o}rper''.
One of the authors (T.K.) thanks Alexander von Humboldt Foundation
for financial support during his stay at University of Hamburg where
the present work has been started.


%
\begin{table}
\begin{tabular}{llllll}
$(M_1,M_2)$ & $\Lambda_c$ & $\nu$ & $E_c$ & $N$ & $\chi^2_{min}$\\ \hline
(6,8) & $0.514 \pm 0.005$ & $1.05 \pm 0.07$ & $4.414 \pm 0.001$ &
$42$ & $\sim 84.3$\\
(8,10) & $0.516 \pm 0.007$ & $1.26 \pm 0.09$ & $4.414 \pm 0.001$ &
$42$ & $\sim 30.0$\\
(10,12) & $0.51 \pm 0.01$ & $1.32 \pm 0.12$ & $4.414 \pm 0.001$ &
$42$ & $\sim 15.9$\\
(12,14) & $0.56 \pm 0.02$ & $1.49 \pm 0.16$ & $4.409 \pm 0.002$ &
$42$ & $\sim 18.7$\\
(14,16) & $0.48 \pm 0.02$ & $1.475 \pm 0.19$ & $4.417 \pm 0.002$ &
$42$ & $\sim 32.4$\\
(12,16) & $0.53 \pm 0.01$ & $1.46 \pm 0.09$ & $4.413 \pm 0.001$ &
$42$ & $\sim 18.6$
\end{tabular}
\caption{Results for $\Lambda_c$, $\nu$, and $E_c$
by the fits using the data for two system-sizes
$M_1$ and $M_2$ for $W=0$. Here $N$ stands for the number of data 
used for the fit, and $\chi^2_{min}$ denotes the minimun value 
of $\chi^2$ obtained by the fit. 
}
\end{table}
\begin{table}
\begin{tabular}{lllllll}
$E$ & $\Lambda_6$ & $\Lambda_8$ & $\Lambda_{10}$ & 
$\Lambda_{12}$ & $\Lambda_{14}$ & $\Lambda_{16}$\\ \hline
$4.3900$ & $0.650 \pm 0.001$ & $0.696 \pm 0.001$ & $0.736 \pm 0.001$ &
$0.777 \pm 0.002$ & $0.808 \pm 0.002$ & $0.843 \pm 0.003$ \\
$4.3925$ & $0.634 \pm 0.001$ & $0.675 \pm 0.001$ & $0.709 \pm 0.001$ &
$0.745 \pm 0.002$ & $0.768 \pm 0.003$ & $0.799 \pm 0.002$ \\
$4.3950$ & $0.620 \pm 0.001$ & $0.654 \pm 0.001$ & $0.686 \pm 0.001$ &
$0.713 \pm 0.002$ & $0.729 \pm 0.002$ & $0.765 \pm 0.003$ \\
$4.3975$ & $0.605 \pm 0.001$ & $0.634 \pm 0.001$ & $0.658 \pm 0.001$ &
$0.683 \pm 0.002$ & $0.696 \pm 0.002$ & $0.721 \pm 0.004$ \\
$4.4000$ & $0.588 \pm 0.001$ & $0.614 \pm 0.001$ & $0.636 \pm 0.001$ &
$0.654 \pm 0.002$ & $0.664 \pm 0.002$ & $0.688 \pm 0.002$ \\
$4.4025$ & $0.574 \pm 0.001$ & $0.595 \pm 0.001$ & $0.609 \pm 0.001$ &
$0.627 \pm 0.002$ & $0.635 \pm 0.002$ & $0.652 \pm 0.005$ \\
$4.4050$ & $0.561 \pm 0.001$ & $0.575 \pm 0.001$ & $0.588 \pm 0.001$ &
$0.600 \pm 0.002$ & $0.603 \pm 0.001$ & $0.613 \pm 0.004$ \\
$4.4075$ & $0.546 \pm 0.001$ & $0.557 \pm 0.001$ & $0.565 \pm 0.001$ &
$0.575 \pm 0.001$ & $0.573 \pm 0.002$ & $0.587 \pm 0.001$ \\
$4.4100$ & $0.530 \pm 0.001$ & $0.540 \pm 0.001$ & $0.546 \pm 0.001$ &
$0.549 \pm 0.001$ & $0.547 \pm 0.002$ & $0.554 \pm 0.003$ \\
$4.4125$ & $0.517 \pm 0.001$ & $0.522 \pm 0.001$ & $0.524 \pm 0.001$ &
$0.526 \pm 0.001$ & $0.522 \pm 0.002$ & $0.529 \pm 0.002$ \\
$4.4150$ & $0.506 \pm 0.001$ & $0.505 \pm 0.001$ & $0.505 \pm 0.001$ &
$0.502 \pm 0.001$ & $0.498 \pm 0.002$ & $0.497 \pm 0.002$ \\
$4.4175$ & $0.494 \pm 0.001$ & $0.492 \pm 0.001$ & $0.484 \pm 0.001$ &
$0.481 \pm 0.001$ & $0.474 \pm 0.002$ & $0.473 \pm 0.001$ \\
$4.4200$ & $0.483 \pm 0.001$ & $0.474 \pm 0.001$ & $0.468 \pm 0.001$ &
$0.461 \pm 0.001$ & $0.453 \pm 0.001$ & $0.452 \pm 0.002$ \\
$4.4225$ & $0.472 \pm 0.001$ & $0.459 \pm 0.001$ & $0.450 \pm 0.001$ &
$0.443 \pm 0.001$ & $0.432 \pm 0.001$ & $0.424 \pm 0.002$ \\
$4.4250$ & $0.457 \pm 0.001$ & $0.445 \pm 0.001$ & $0.433 \pm 0.001$ &
$0.423 \pm 0.001$ & $0.411 \pm 0.001$ & $0.404 \pm 0.002$ \\
$4.4275$ & $0.448 \pm 0.001$ & $0.431 \pm 0.001$ & $0.417 \pm 0.001$ &
$0.405 \pm 0.001$ & $0.393 \pm 0.002$ & $0.383 \pm 0.001$ \\
$4.4300$ & $0.438 \pm 0.001$ & $0.418 \pm 0.001$ & $0.401 \pm 0.001$ &
$0.387 \pm 0.001$ & $0.373 \pm 0.001$ & $0.364 \pm 0.001$ \\
$4.4325$ & $0.428 \pm 0.001$ & $0.405 \pm 0.001$ & $0.387 \pm 0.001$ &
$0.371 \pm 0.001$ & $0.355 \pm 0.001$ & $0.345 \pm 0.001$ \\
$4.4350$ & $0.420 \pm 0.001$ & $0.391 \pm 0.001$ & $0.374 \pm 0.001$ &
$0.356 \pm 0.001$ & $0.339 \pm 0.001$ & $0.328 \pm 0.001$ \\
$4.4375$ & $0.409 \pm 0.001$ & $0.382 \pm 0.001$ & $0.359 \pm 0.001$ &
$0.342 \pm 0.001$ & $0.323 \pm 0.001$ & $0.313 \pm 0.001$ \\
$4.4400$ & $0.402 \pm 0.001$ & $0.369 \pm 0.001$ & $0.347 \pm 0.001$ &
$0.328 \pm 0.001$ & $0.308 \pm 0.002$ & $0.298 \pm 0.001$
\end{tabular}
\caption{List of numerical data of $\Lambda_M$ 
presented in figure 1 for $W=0$.}
\end{table}
\begin{figure}
\hspace*{1cm}\epsfxsize=10cm \epsfbox{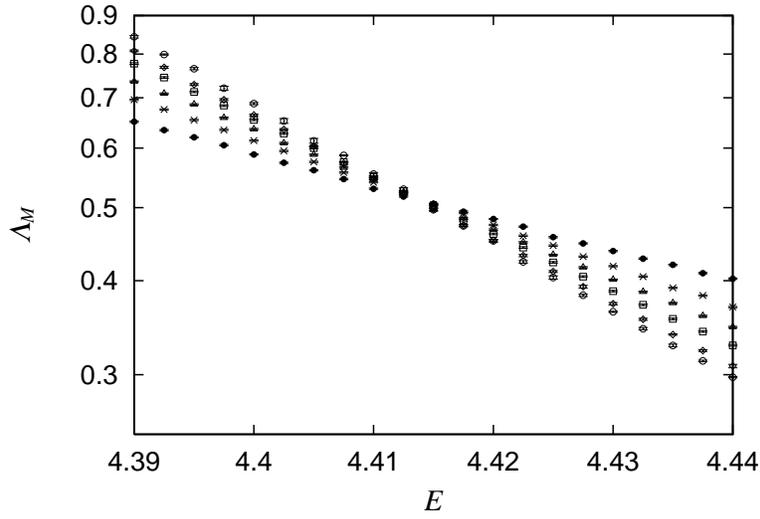}
\caption{Renormalized localization lengths for $W=0$ are 
presented in log-scale as a function of energy.
The dots, the crosses, the triangles,
the squares, the diamonds, the circles correspond to $M=6$, $8$, $10$,
$12$, $14$, and $16$, respectively.}
\end{figure}
\begin{figure}
\hspace*{2cm}\epsfxsize=10cm \epsfbox{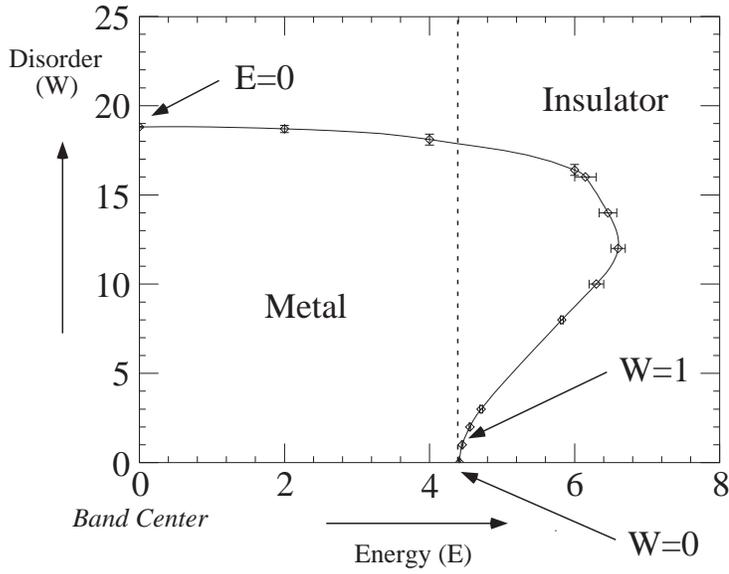}
\caption{Mobility edge trajectory for the 
3D system in the random magnetic field. The re-entrant phaenomenon 
is clearly seen.}
\end{figure}
\begin{figure}
\hspace*{2cm}\epsfxsize=10cm \epsfbox{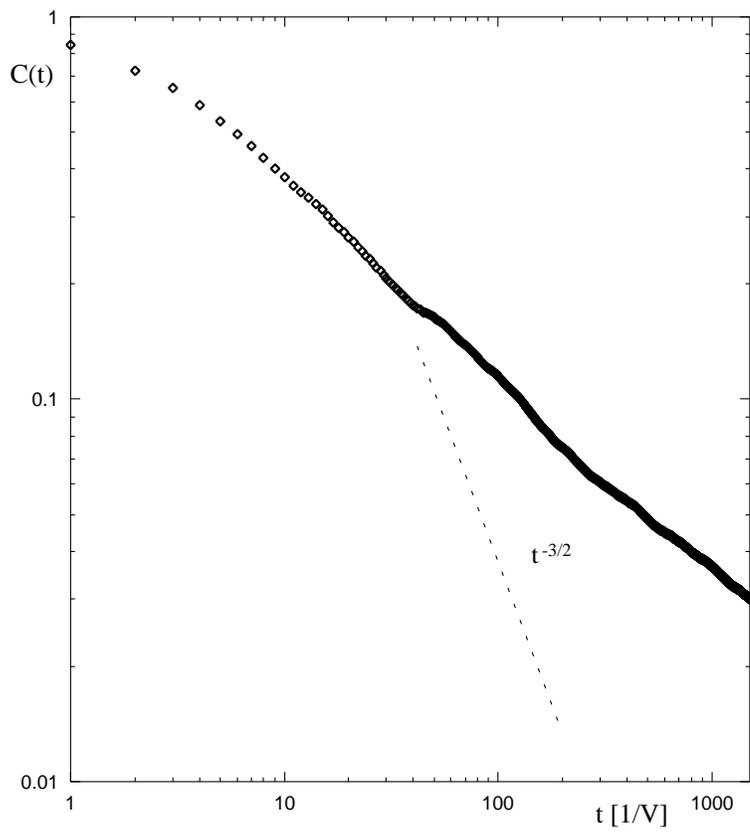}
\caption{The autocorrelation function $C(t)$ as a function of $t$.
Dashed lines represent the decay $\propto t^{-3/2}$.}
\end{figure}

\begin{references}
\bibitem{Anderson} P.W. Anderson, Phys. Rev. {\bf 109} (1958) 1492 
\bibitem{Wegner} F. Wegner, Z. Phys. {\bf B25} (1976) 327; {\bf B35} (1979) 207
\bibitem{Hikami} S. Hikami, Phys. Rev. {\bf B24} (1981) 2671
\bibitem{LR} P.A. Lee and T.V. Ramakrishnan, Rev. Mod. Phys. {\bf 57}
             (1985) 287 
\bibitem{KM} B. Kramer and A. MacKinnon, Rep. Prog. Phys. {\bf 56}
             (1993) 1469 
\bibitem{Dyson} F.J. Dyson, J. Math. Phys. {\bf 3}, (1962) 140;
                {\bf 3} (1962) 157; {\bf 3} (1962) 166
\bibitem{MK} A. MacKinnon and B. Kramer, Phys. Rev. Lett. {\bf 47} 
             (1981) 1546; Z. Phys. B {\bf 53} (1983) 1
\bibitem{KBMS} B. Kramer, K. Broderix, A. MacKinnon and M. Schreiber,
              Physica A {\bf 167} (1990) 163
\bibitem{MacKinnon} A. MacKinnon, J. Phys.: Condens. Matter {\bf 6}
              (1994) 2511
\bibitem{SO98} K. Slevin and T. Ohtsuki, Phys. Rev. Lett. {\bf 82}
             (1999) 382
\bibitem{SO} K. Slevin and T. Ohtsuki, Phys. Rev. Lett. {\bf 78}
             (1997) 4083
\bibitem{SK} M. Schreiber and B. Kramer, 
             in {\it Anderson Transition} edited
             by T.Ando and H.Fukuyama, 92 (Springer-Verlag 1988).
\bibitem{Hajdu} {\it Introduction to the Theory of the Integer
Quantum Hall Effect}, J. Hajdu {\it et al.},
             (VCH, 1994)
\bibitem{Huckestein} B. Huckestein, Rev. Mod. Phys. {\bf 67} (1994) 357
\bibitem{OOK} T. Ohtsuki, Y.Ono and B. Kramer, J. Phys. Soc. Jpn.
             {\bf 63} (1994) 685
              \label{OOK}
\bibitem{AI} B. Altshuler and L.B. Ioffe, Phys. Rev. Lett. {\bf 69}
(1992) 2979
\bibitem{Furusaki} A. Furusaki, Phys. Rev. Lett. 82 (1999) 604.
\bibitem{HLR} B.I. Halperin, P.A. Lee, and N. Read,Phys. Rev. {\bf B47}
(1993) 7312
\bibitem{OKO} T. Ohtsuki, B. Kramer and Y. Ono, J. Phys. Soc. Jpn.
              {\bf 62} (1993) 224
\bibitem{HKO} M. Henneke, B. Kramer and T. Ohtsuki, Euro. Phys. Lett.
              {\bf 27} (1994) 389
\bibitem{CD} J.T. Chalker and A. Dohmen, Phys. Rev. Lett. {\bf 75} 
(1995) 4496
\bibitem{KOH} B. Kramer, T. Ohtsuki and M. Henneke, in {\it
              Quantum Dynamics of Submicron Structures}, edited by
              H.A. Cerdeira et al., 21
              (Kluwer Academic Publishers, 1995)
\bibitem{For} W.H. Press {\it et al.}, 
              Numerical Recipes (Cambridge University
              Press, 1986)
\bibitem{KKO} T. Kawarabayashi, B. Kramer, and T.Ohtsuki, Phys. Rev.
              {\bf B57} (1998) 11842
              \label{KKO}
\bibitem{KKO2} T. Kawarabayashi, B. Kramer, and T. Ohtsuki, J. Phys.
              : Condens. Matter {\bf 10} (1998) 11547 
\bibitem{BSK} B. Bulka, M. Schreiber, and B. Kramer, Z. Phys. 
              {\bf B66} (1987) 21
\bibitem{DBZK} T. Dr\"{o}se,M. Batsch,I. Zharekeshev, and B. Kramer,
              Phys. Rev. {\bf B57}(1998) 37
\bibitem{fractal} H. Aoki, J. Phys. C {\bf 16} (1983) L205;
C.M. Soukoulis and E.N. Economou, Phys. Rev. Lett. {\bf 52} (1984)
565;
M. Schreiber, Phys. Rev. B {\bf 31} (1985) 6146;
Y. Ono, B. Kramer and T. Ohtsuki, J. Phys. Soc. Jpn. {\bf 58}
(1989) 1705;
M. Schreiber and H. Grussbach, Phys. Rev. Lett. {\bf 67} (1991) 607;
B. Pook and M. Janssen, Z. Phys. B {\bf 82} (1991) 295
\bibitem{shapiro} B. Shapiro, Phys. Rev. Lett. {\bf 65} (1990) 1510;
A. Cohen and B. Shapiro, { Int. J. Mod. Phys.} {\bf B6} (1992)
1243
\bibitem{markos} P. Markos and B. Kramer { Phil. Mag.} {\bf B68} (1993) 357
\bibitem{SSSLS} B.I. Shklovskii, B. Shapiro, B.R. Sears, P. Lambrianides
and H.B. Shore, Phys. Rev. {\bf B47} (1993) 11487 
\bibitem{OO} Y. Ono and T. Ohtsuki, J. Phys. Soc. Jpn. {\bf 62}
 (1993) 3813
\bibitem{HS_3DO} E. Hofstetter and M. Schreiber, Phys. Rev. {\bf B48}
(1993) 16979; {\bf B49} (1994) 14726
\bibitem{Evangelou_3DO} S.N. Evangelou, Phys. Rev. {\bf B49} (1994) 16805
\bibitem{ZK} I. Kh. Zharekeshev and B. Kramer,
 Jpn. J. Appl. Phys. {\bf 34} (1995) 4361; 
 Phys. Rev. {\bf B51} (1995) 17239;
 Phys. Rev. Lett {\bf 79} (1997) 717
\bibitem{BSZK} M. Batsch, L. Schweitzer,  I. Kh. Zharekeshev and B. Kramer,
Phys. Rev. Lett. {\bf 77}  (1996)  1552
\bibitem{SZ} L. Schweitzer and  I. Kh. Zharekeshev,
J. Phys.: Condens. Matter {\bf 9} (1997)  L377
\bibitem{KOSO}T. Kawarabayashi, T. Ohtsuki, K. Slevin and Y. Ono,
Phys. Rev. Lett. {\bf 77} (1996) 3593
\bibitem{Suzuki} M. Suzuki, { Phys. Lett.} {\bf A146} (1990) 319;
{ J. Math. Phys.} {\bf 32} (1991) 400; J. Phys. Soc. Jpn.
{\bf 61} (1992) 3015; Commun. Math. Phys. {\bf 163} (1994)  491
\bibitem{KO} T. Kawarabayashi and T. Ohtsuki, Phys. Rev. {\bf B51}
(1995) 10897
\bibitem{shapiro2} B. Shapiro: in
{\it Percolation structures and Process}, ed. G. Deutscher {\it et al.}
Ann. Isr. Phys. Soc. (1983) 367
\bibitem{OK} T. Ohtsuki and T. Kawarabayashi, J. Phys. Soc. Jpn.{\bf
66} (1997) 314
\bibitem{BHS} T. Brandes, B. Huckestein, and L. Schweitzer, Ann.
Physik 5 (1996) 633
\bibitem{HK} B. Huckestein and R. Klesse, Phil. Mag. {\bf B77} (1998)
1181
\end{references}
\end{document}